\documentclass[prl,twocolumn,showpacs,preprintnumbers,amsmath,amssymb]{revtex4}

\usepackage{graphicx}
\usepackage{rotating}
\usepackage{dcolumn}
\usepackage{bm}

\begin{document}

\preprint{cond-mat/somewhere}

\title{`St\"uckelberg interferometry' with ultracold molecules} 

\author{M. Mark,$^{1}$  T. Kraemer,$^{1}$ P. Waldburger,$^{1}$ J. Herbig,$^{1}$ C. Chin,$^{1,3}$
H.-C. N\"{a}gerl,$^{1}$ R. Grimm$^{1,2}$}
\affiliation{ $^1$Institut f\"{u}r Experimentalphysik und
Forschungszentrum f\"{u}r Quantenphysik,
Universit\"{a}t Innsbruck,  6020 Innsbruck, Austria \\
$^2$Institut f\"{u}r Quantenoptik und Quanteninformation,
\"{O}sterreichische Akademie der Wissenschaften, 6020 Innsbruck,
Austria \\ $^3$Physics Department and James Franck Institute,
University of Chicago, Chicago, IL 60637, USA}

\date{\today}

\begin{abstract}
We report on the realization of a time-domain `St\"uckelberg
interferometer', which is based on the internal state structure of
ultracold Fesh\-bach molecules. Two subsequent passages through a
weak avoided crossing between two different orbital angular
momentum states in combination with a variable hold time lead to
high-contrast population oscillations. This allows for a precise
determination of the energy difference between the two molecular
states. We demonstrate a high degree of control over the
interferometer dynamics. The interferometric scheme provides new
possibilities for precision measurements with ultracold molecules.
\end{abstract}


\maketitle

The creation of molecules on Feshbach resonances in atomic quantum
gases has opened up a new chapter in the field of ultracold matter
\cite{Koehler2007}. Molecular quantum gases are now readily
available in the lab for various applications. Prominent examples
are given by the creation of strongly interacting many-body
systems based on molecular Bose-Einstein condensates
\cite{Fermi-Systems}, experiments on few-body collision physics
\cite{Chin2005}, the realization of molecular matter-wave optics
\cite{Abo-Shaeer2005}, and by the demonstration of exotic pairs in
optical lattices \cite{Winkler2006}. Recent experimental progress
has shown that full control over all degress of freedom can be
expected for such molecules
\cite{Thalhammer2006,Volz2006,Winkler2007}. Ultracold molecular
samples with very low thermal spread and long interaction times
could greatly increase the sensitivity in measurements of
fundamental physical properties such as the existence of an
electron dipole moment \cite{Hudson2002} and a possible
time-variation of the fine-structure constant
\cite{Hudson2006,Chin2006}.

Most of today's most accurate and precise measurements rely on
interferometric techniques applied to ultracold atomic systems.
For example, long coherence times in atomic fountains or in
optical lattices allow ultraprecise frequency metrology
\cite{Bize2005,Boyd2007}. Molecules, given their rich internal
structure, greatly extend the scope of possible precision
measurements. Molecular clocks, for example, may provide novel
access to fundamental constants and interaction effects, different
from atomic clocks. The fast progress in preparing cold molecular
samples thus opens up fascinating perspectives for precision
interferometry. Recently, the technique of Stark deceleration has
allowed a demonstration of Ramsey interferometry with a cold and
slowed molecular beam \cite{Hudson2006}. Ultracold trapped
molecular ensembles are expected to further enhance the range of
possible measurements.

In this Letter, we report on the realization of an
inter\-nal-state interferometer with ultracold Cs$_{2}$ molecules.
A weak avoided crossing is used as a `beam splitter' for molecular
states as a result of partial Landau-Zener tunneling when it is
traversed by means of an appropriately chosen magnetic field ramp.
Using the avoided crossing twice, first for splitting, and then
for recombination of molecular states, leads to the well-known
`St\"uckelberg oscillations' \cite{Stueckelberg1932}. We thus call
our scheme a `St\"uckelberg interferometer'. Our realization of
this interferometer allows full control over the interferometer
dynamics. In particular, the hold time between splitting and
recombination can be freely chosen. In analogy to the well-known
Ramsey interferometer \cite{Ramsey1956} the acquired
interferometer phase is mapped onto the relative populations of
the two output states that can be well discriminated upon
molecular dissociation. To demonstrate the performance of the
St\"uckelberg interferometer we use it for precision molecular
spectroscopy to determine the position and coupling strength of
the avoided crossing.

\begin{figure}
\includegraphics[width=0.48\textwidth]{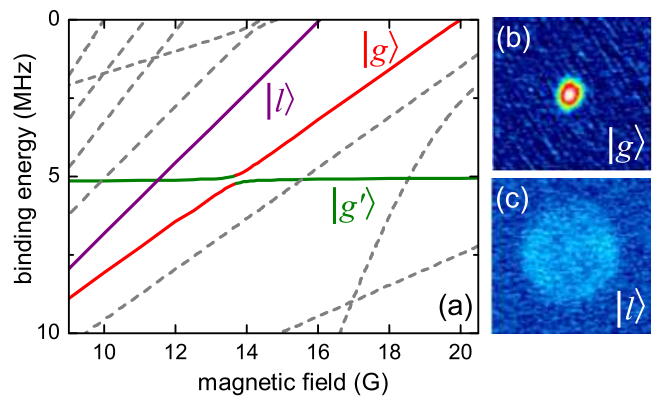}
\caption{(color online). (a) Molecular energy structure below the
dissociation threshold showing all molecular states up $
\ell\!=\!8 $. The relevant states for the present experiment
(solid lines) are labeled $|g\rangle$, $|g'\rangle$, and
$|l\rangle$. Molecules in state $|g \rangle$ or $|l\rangle$ are
detected upon dissociation as shown in (b) and (c). The crossing
used for the interferometer is the one between $|g'\rangle$ and
$|l\rangle$ near $11.4$\,G. Initially, ultracold molecules are
generated in state $|g\rangle$ on the Feshbach resonance at
$19.8$G.} \label{fig_structure}
\end{figure}

The energy structure of weakly bound Cs$_2$ dimers in the relevant
range of low magnetic field strength is shown in
Fig.~\ref{fig_structure} \cite{Chin2004}. Zero binding energy
corresponds to the threshold of dissociation into two free Cs
atoms in the lowest hyperfine sublevel $|F\!=\!3,
m_F\!=\!3\rangle$ and thus to the zero-energy collision limit of
two atoms. The states relevant for this work are labelled by
$|g\rangle$, $|g'\rangle$, and $|l\rangle$ \cite{molstates}. While
$|g\rangle$ and $|g'\rangle$ are $g$-wave states with orbital
angular momentum $\ell\!=\!4$, the state $|l\rangle$ is an
$l$-wave state with a high orbital angular momentum of
$\ell\!=\!8$ \cite{Eite}. Coupling with $\Delta \ell\!=\!4$
between s-wave atoms and $g$-wave molecules and between $g$- and
$l$-wave states is a result of the strong indirect spin-spin
interaction between two Cs atoms \cite{Chin2004}.

The starting point for our experiments is a Bose-Einstein
condensate (BEC) with $\sim\!2\times 10^5$ Cs atoms in the
$|F\!=\!3, m_F\!=\!3\rangle$ ground state confined in a
crossed-beam dipole trap generated by a broad-band fiber laser
with a wavelength near 1064\,nm \cite{Weber2003,Kraemer2004}. The
BEC allows us to efficiently produce molecules on a narrow
Fesh\-bach resonance at $19.84$~G \cite{Herbig2003} in an
optimized scheme as described in Ref.~\cite{Mark2005}. With an
efficiency of typically 25\% we produce a pure molecular ensemble
with up to $2.5 \times 10^{4}$ ultracold molecules all in state
$|g\rangle$, initially close to quantum degeneracy
\cite{Herbig2003}. The following experiments are performed on the
molecular ensemble in free fall. During the initial expansion to a
$1/e$-diameter of about $28 \, \mu$m along the radial and about
$46 \, \mu$m along the axial direction the peak density is reduced
to $1 \times 10^{11}$cm$^{-3}$ so that molecule-molecule
interactions \cite{Chin2005} can be neglected on the timescale of
the experiment.

\begin{figure}
\includegraphics[width=0.48\textwidth]{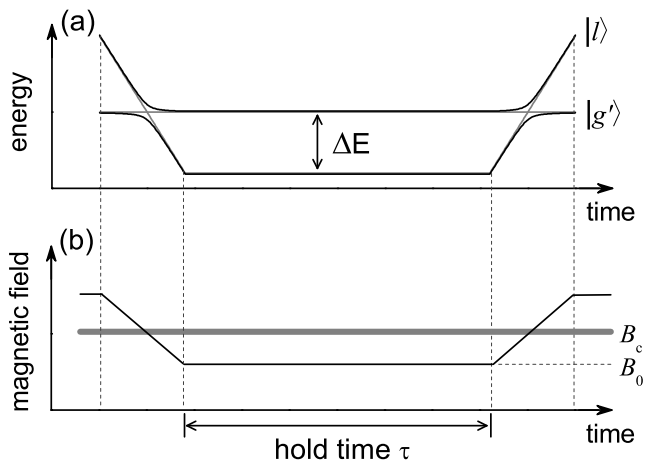}
\caption{(a) Scheme of the `St\"uckelberg interferometer'. By
ramping the magnetic field over the avoided crossing at $ B_c $ at
a rate near the critical ramp rate $ R_c $ the population in the
initial molecular state is coherently split. $\Delta E $ is the
binding energy difference at the given hold field $B_0$. After the
hold time $\tau$ a reverse ramp coherently recombines the two
populations. The populations in the two `output ports' are then
determined as a function of acquired phase difference $ \phi
\propto \Delta E \times \tau$. (b) Corresponding magnetic field
ramp.} \label{fig_scheme}
\end{figure}

The molecules can now be transferred to any one of the molecular
states shown in Fig.~\ref{fig_structure} with near 100\%
efficiency by controlled `jumping' or adiabatic following at the
various crossings \cite{michael_dimers}. When the magnetic field
strength is decreased, the molecules first encounter the crossing
at $13.6$\,G. At all ramp rates used in our present experiments
the passage through this crossing takes place in a fully adiabatic
way. The molecules are thus transferred from $|g\rangle$ to
$|g'\rangle$ along the upper branch of the crossing. They then
encounter the next crossing at a magnetic field of $B_c \approx
11.4\,$G. We accidentally found this weak crossing in our previous
magnetic moment measurements \cite{Chin2005,michael_dimers}. This
allowed the identification of the $l$-wave state $|l\rangle$
\cite{Eite}.

This crossing between $|g'\rangle$ and $|l\rangle$ plays a central
role in the present experiment. It can be used as a tunable `beam
splitter', which allows adiabatic transfer, coherent splitting, as
will be shown below, or diabatic transfer for the molecular states
involved, depending on the chosen magnetic ramp rate near the
crossing. We find that a critical ramp rate of $ R_c \approx 14
$\,G/ms leads to a 50/50-splitting into $|g'\rangle$ and
$|l\rangle$ \cite{michael_dimers}. Using the well-known
Landau-Zener formula and an estimate for the difference in
magnetic moment for states $|g'\rangle$ and $|l\rangle$
\cite{Eite} we determine the coupling strength $ V $ between
$|g'\rangle$ and $|l\rangle$ to $\sim\! h \! \times \! 15$ kHz.

\begin{figure*}
\includegraphics[width=1.0\textwidth]{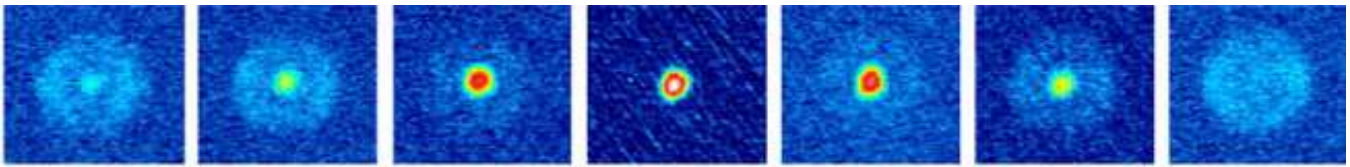}
\caption{(color online). Series of dissociation patterns showing
about one oscillation period with $\Delta E/h=155$\,kHz at a hold
field of 11.19\,G. From one picture to the next the hold time $
\tau$ is increased by 1\,$\mu$s. The first and the last of the
absorption images mainly show dissociation of $l$-wave molecules,
whereas the central image shows predominant dissociation of
$g$-wave molecules.} \label{fig_pics}
\end{figure*}

We state-selectively detect the molecules by ramping up the
magnetic field to bring the molecules above the threshold for
dissociation. There the quasi-bound molecules decay into the
atomic scattering continuum. For state $|g\rangle$ dissociation is
observed for magnetic fields above the 19.84\,G position of the
corresponding Feshbach resonance. Fig.~\ref{fig_structure} (b)
shows a typical absorption image of the resulting atom cloud
\cite{Herbig2003}. For state $|l\rangle$ dissociation is observed
above 16.5\,G. The molecular states can thus be easily
discriminated by the different magnetic field values needed for
dissociation. Moreover, the expansion pattern is qualitatively
different from the one connected to state $|g\rangle$. The
absorption image in Fig.~\ref{fig_structure} (c) shows an
expanding `bubble' of atoms with a relatively large kinetic energy
of about $ k_B \! \times \! 20 \, \mu$K per atom. Here, $ k_B $ is
Boltzmann's constant. We find that significant dissociation occurs
only when the state $|l\rangle$ couples to a quasi-bound $g$-wave
state about $h \! \times \! 0.7$ MHz above threshold
\cite{Steven}. The resulting bubble is not fully spherically
symmetric, which indicates higher partial-wave contributions
\cite{Duerr2004}. The different absorption patterns allow us to
clearly distinguish between the two different dissociation
channels in a single absorption picture when the magnetic field is
ramped up to $\sim\! 22$ G. These dissociation channels serve as
the interferometer `output ports'.

The interferometer is based on two subsequent passages through the
crossing following the scheme illustrated in
Fig.~\ref{fig_scheme}. For an initial magnetic field above the
crossing a downward magnetic-field ramp brings the initial
molecular state into a coherent superposition of $|g'\rangle$ and
$|l\rangle$. After the ramp the field is kept constant at a hold
field $B_0$ below the crossing for a variable hold time $\tau$. A
differential phase $ \phi $ is then accumulated between the two
components, which linearly increases with the product of the
binding energy difference $\Delta E$ and the hold time $\tau$. The
magnetic field is then ramped back up, and the second passage
creates a new superposition state depending on $ \phi $. For a
50/50-splitting ratio, this can lead to complete destructive or
constructive interference in the two output ports and thus to
high-contrast fringes as a function of $\tau$ or $\Delta E$. These
fringes, resulting from two passages through the same crossing,
are analogous to the well-known St\"uckelberg oscillations in
collision physics \cite{Stueckelberg1932,Nikitin1984} or in the
physics with Rydberg atoms \cite{Baruch1992,Yoakum1992}. Note that
our realization of a `St\"uckelberg interferometer' gives full
control over the interferometer dynamics by appropriate choice of
ramp rates and magnetic offset fields.

A typical ramp sequence, as shown in Fig.~\ref{fig_scheme} (b),
starts with a sample of $|g'\rangle$ molecules at a magnetic field
of 11.6\,G about 250\,mG above the crossing. At the critical ramp
rate $ R_c $ we ramp the magnetic field within about $ 50 \, \mu$s
to a hold field $B_0$ below the crossing. After the variable hold
time $\tau$ we reverse the ramp and transverse the crossing a
second time at the critical ramp rate. The output of the
interferometer is detected by rapidly ramping the magnetic field
up to 22\,G and by imaging the pattern of dissociating $|l\rangle$
and $|g\rangle$ molecules.

\begin{figure}
\includegraphics[width=0.48\textwidth]{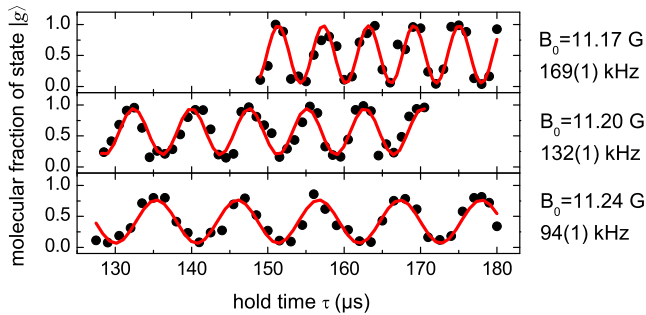}
\caption{Interferometer fringes for magnetic hold fields $ B_0$
below the crossing of $|g'\rangle$ and $|l\rangle$. The $g$-wave
molecular fraction is plotted as a function of the hold time
$\tau$. Sinusoidal fits give the oscillation frequency as
indicated.} \label{fig_oscill}
\end{figure}


For one period of oscillation the dependence of the dissociation
pattern on the hold time $\tau$ is demonstrated by the series of
absorption images shown in Fig.~\ref{fig_pics}. The hold time is
increased in steps of 1\,$\mu$s while the entire preparation,
ramping, and detection procedure is repeated for each experimental
cycle, lasting about 20 s. The molecular population oscillates
from being predominantly $l$-wave to predominatly $g$-wave and
back. For a quantitative analysis of the molecular population in
each output port we fit the images with a simple model function
\cite{waldburger} and extract the fraction of molecules in each of
the two output ports. Fig.~\ref{fig_oscill} shows the $g$-wave
molecular population as a function of hold time $\tau$ for various
hold fields $B_0$ corresponding to different $ \Delta E$. The
existence of these St\"uckelberg oscillations confirms that
coherence is preserved by the molecular beam splitter. Their high
contrast ratio shows that near 50/50-splitting is achieved.
Sinusoidal fits to the data allow for an accurate determination of
the oscillation frequency and hence of $\Delta E$.



\begin{figure}
\includegraphics[width=0.45\textwidth]{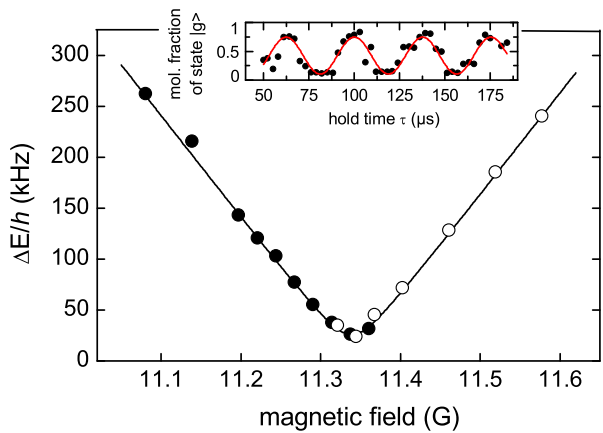}
\caption{Interferometrically measured binding energy difference
$\Delta E$ in the region of the crossing between states
$|g'\rangle$ and $|l\rangle$ as a function of magnetic field.
Solid circles: Standard ramp sequence of the interferometer. Open
circles: Inverted scheme for field values above the crossing. The
one-sigma statistical error from the sinusoidal fit is less than
the size of the symbols. The solid curve is a hyperbolic fit to
the experimental data. Inset:  Oscillation at 26.6(3) kHz for a
hold field $B_0 = 11.34$ G right on the crossing.} \label{fig5}
\end{figure}

Fig.~\ref{fig5} shows $\Delta E$ as a function of magnetic field
strength near the avoided crossing. For magnetic fields below the
crossing we obtain $\Delta E$ as described before. For magnetic
fields above the crossing, we invert the interferometric scheme.
Molecules are first transferred from $|g'\rangle$ into $|l\rangle$
using a slow adiabatic ramp. The field is then ramped up above the
crossing with a rate near $ R_c $, kept constant for the variable
time $\tau$ at the hold field $B_0$ and then ramped down to close
the interferometer. An adiabatic ramp through the crossing maps
population in $|g'\rangle$ onto $|l\rangle$ and vice versa.
Detection then proceeds as before.

For both realizations of the interferometer we obtain
high-contrast fringes even when it is not operated in the
Landau-Zener regime and the fast ramps are stopped right at the
crossing (see inset to Fig.~\ref{fig5}). This allows us to measure
$\Delta E$ in the crossing region. A fit to the data with a
hyperbolic fit function according to the standard Landau-Zener
model yields $ B_c=11.339(1)$ G for the position of the crossing,
$\Delta \mu = 0.730(6) \, \mu_B$ for the difference in magnetic
moment of the two states involved ($\mu_B$ is Bohr's magneton),
and $V=h\!\times\!14(1)$ kHz for the coupling strength. While the
measured $\Delta \mu$ agrees reasonably well with the result from
an advanced theoretical model of the Cs$_2$ dimer \cite{Eite}, $
B_c $ and $V$ cannot be obtained from these calculations with
sufficient accuracy.


The present interferometer allows us to observe up to 100
oscillations at 200 kHz. Shot-to-shot fluctuations increasingly
scramble the phase of the oscillations for longer hold times until
the phase appears to be fully randomized while large amplitude
variations for the molecular populations persist. The peak-to-peak
amplitude of these fluctuations decays slowly and is still 50\% of
the initial contrast after 1 ms. We attribute this phase
scrambling to magnetic field noise that causes shot-to-shot
variations of $\Delta E$, the same, however, for each molecule.
The large amplitude of these fluctuations indicates that phase
coherence is preserved within the molecular sample. We attribute
the gradual loss of peak-to-peak amplitude to spatial magnetic
field inhomogeneities. We expect that straightforward technical
improvements regarding the magnetic field stability and
homogeneity and applying the interferometer to trapped molecular
samples will allow us extend the hold times far into the
millisecond range. It will then be possible to measure ultraweak
crossings with coupling strengths well below $h\!\times\! 1$ kHz.

We have demonstrated a molecular St\"uckelberg interferometer with
full control over the interferometer dynamics. The interferometer
can be used as a spectros\-copic tool as it allows precise
measurements of binding energy differences of molecular states. In
particular, the technique can be employed to measure feeble
interactions between molecular states, such as parity
non-conserving interactions \cite{Commins1999}. The sensitivity to
detect ultraweak level crossings, combined with long storage times
in optical molecule traps \cite{Chin2005} or lattices
\cite{Winkler2006,Thalhammer2006,Volz2006}, may allow to detect
interaction phenomena on the $h\!\times\! 1$ Hz scale. In view of
the rapid progress in various preparation methods for cold
molecular samples, new tools for precision measurements on
molecular samples, such as our St\"uckelberg interferometer, will
open up exciting avenues for future research.

We thank E. Tiesinga for discussions and for theoretical support.
We acknowledge financial support by the Austrian Science Fund
(FWF) within SFB 15 (project part 16) and by the EU within the
Cold Molecules TMR Network under contract No.\ HPRN-CT-2002-00290.
M.M. and C.C. acknowledge support by DOC [PhD-Program of the
Austrian Academy of Science] and the FWF Lise-Meitner program,
respectively.

\end{document}